\documentclass[12pt]{article}

\usepackage{epsfig}

\newcommand{\be}{\begin{equation}}
\newcommand{\ee}{\end{equation}}
\newcommand{\ba}{\begin{eqnarray}}
\newcommand{\ea}{\end{eqnarray}}
\newcommand{\no}{\nonumber\\}

\newcommand{\lesssim}{\:\mbox{\raisebox{-3pt}{$\stackrel%
{\displaystyle <}{\sim}$}}\:}

\begin{document}

\title{\normalsize \hfill UWThPh-2007-27 \\[8mm]
\LARGE $A_4$ model for the quark mass matrices}

\author{
Lu\'\i s Lavoura$^{(1)}$\thanks{E-mail: balio@cftp.ist.utl.pt}
\ and
Helmut K\"uhb\"ock$^{(2)}$\thanks{E-mail: helmut.kuehboeck@gmx.at}
\\*[5mm]
$^{(1)} \!$ \small
Centro de F\'\i sica Te\'orica de Part\'\i culas \\
\small
Instituto Superior T\'ecnico, 1049-001 Lisboa, Portugal
\\*[2mm]
$^{(2)} \!$ \small
Fakult\"at f\"ur Physik, Universit\"at Wien \\
\small
Boltzmanngasse 5, 1090 Wien, Austria
\\*[2mm]
}

\date{28 January 2008}

\maketitle

\begin{abstract}
We propose a model for the quark masses and mixings
based on an $A_4$ family symmetry.
Three scalar $SU(2)$ doublets form a triplet of $A_4$.
The three left-handed-quark $SU(2)$ doublets
are also united in a triplet of $A_4$.
The right-handed quarks are singlets of $A_4$.
The $A_4$-symmetric scalar potential leads to a vacuum
in which two of the three scalar $SU(2)$ doublets
have expectation values with equal moduli.
Our model makes an excellent fit
of the observed $\left| V_{ub} / V_{cb} \right|$.
The symmetry $CP$ is respected
in the charged gauge interactions of the quarks.
\end{abstract}

\section{Introduction}

In the standard $SU(2) \times U(1)$ gauge model
for the electroweak interactions,
the Yukawa couplings of the fermions
to the unique scalar $SU(2)$ doublet of the model
are completely arbitrary---as a matter of fact,
those couplings make up almost all the free parameters
of the Standard Model (SM).
As a consequence of this fact,
the SM leaves the quark masses and mixings unpredicted;
although the model accounts for the existence
of quark masses and mixings,
their actual values remain arbitrary in the context of the SM.

Several more complex models
have tried to overcome this shortcoming of the SM.
However,
most of those models are in reality {\it Ans\"atze}:
instead of {\em deriving} the structure of the Yukawa couplings
from some underlying symmetry of a self-consistent gauge theory,
they simply {\em assume} the Yukawa couplings
to have some aesthetically appealing pattern or texture.
A model should instead rely on some flavour (family) symmetry.

If the flavour symmetry is Abelian,
then all its irreducible representations (irreps) are one-dimensional
and the symmetry can at most force some Yukawa couplings
to vanish.\footnote{The converse of this statement also holds:
any pattern of vanishing Yukawa couplings may be enforced by
an Abelian flavour symmetry
with an adequate spectrum of scalars~\cite{zeros}.}
A non-Abelian flavour symmetry
can also force non-vanishing Yukawa couplings
to be interrelated among themselves
through definite Clebsch--Gordan factors.
Since there are three families of quarks,
a most desirable non-Abelian flavour symmetry
ought to have three-dimensional irreps,
in order to achieve full unification of the three generations and,
hence,
to achieve a minimal number of independent Yukawa couplings.
The smallest discrete group
with a three-dimensional irrep is $A_4$,\footnote{A useful list
of all the discrete non-Abelian groups
with 31 or less group elements is provided in~\cite{kephart}.}
the group of the even permutations of four objects.
The group $A_4$ has 12 group elements,
one triplet irrep $\bf{3}$ and three inequivalent singlet irreps $\bf{1}$,
$\bf{1^\prime}$,
and $\bf{1^{\prime\prime}}$
(the $\bf{1}$ is the trivial representation,
the irreps $\bf{1^\prime}$ and $\bf{1^{\prime\prime}}$
are complex-conjugate of each other).
This group has,
in the last few years,
been used in many models
for the lepton masses and mixings~\cite{leptons}.
It has also been used in models for the quark sector~\cite{quarks},
or for both quarks and leptons simultaneously~\cite{both}.

In this paper we suggest a model for the Yukawa couplings of the quarks
based on an $A_4$ family symmetry.
Our model has three scalar gauge-$SU(2)$ doublets
united in a $\bf{3}$ of $A_4$.
The left-handed-quark $SU(2)$ doublets
are also united in a $\bf{3}$ of $A_4$.
In each electric-charge sector,
the three right-handed quarks are in a $\bf{1}
\oplus \bf{1^\prime} \oplus \bf{1^{\prime\prime}}$ of $A_4$.
Thus,
our model achieves a high degree of simplicity and,
even,
uniqueness,
because it treats the three families of quarks in the same way
and it treats both electric-charge sectors in the same way.
Furthermore,
our model does not require $A_4$ to be broken anywhere in the Lagrangian,
not even through soft terms---the breaking of $A_4$ is solely spontaneous.
This,
too,
adds to the simplicity of the model.

Surprisingly,
our model is able to predict the mixing parameter
$\left| V_{ub} / V_{cb} \right|$
(where $V$ is the quark mixing,
or CKM,
matrix)
fully right:
it predicts $\left| V_{ub} / V_{cb} \right| \approx
0.088$,
in agreement with the usual averages
of the various phenomenological analyses.
On the other hand,
our model also leads to a null violation of the discrete symmetry $CP$
in the charged gauge interactions of the quarks;
thus,
the observed $CP$ violation,
for instance in $K^0$--$\bar K^0$ mixing,
or in $B^0_d$ decays,
should in the context of our model
be explained through scalar-mediated interactions,
including flavour-changing neutral Yukawa interactions.

The plan of our paper is the following.
In section~\ref{yukawas} we derive the form
of the quark Yukawa-coupling matrices.
In section~\ref{potential} we study the scalar potential
and the ensuing vacuum.
In section~\ref{matrices} we write down the quark mass matrices
and demonstrate that in our model there is no $CP$ violation
in the CKM matrix.
In section~\ref{chi2}
we explain the method that we used in the numerical analysis
and give some fits and results.
A short summary is provided in section~\ref{summary}.

\section{The Yukawa couplings}
\label{yukawas}

The gauge symmetry of the model is $SU(2) \times U(1)$.
There are three scalar $SU(2)$ doublets $\phi_j$ ($j = 1, 2, 3$)
with hypercharge $1/2$.
They form a triplet $\bf{3}$ of the flavour symmetry $A_4$.
There are three left-handed-quark $SU(2)$ doublets $Q_{Lj}$
with hypercharge $1/6$.
They are united in another $\bf{3}$ of $A_4$.
There are
three right-handed-quark $SU(2)$ singlets $n_{Rj}$
with hypercharge $-1/3$
and three right-handed-quark $SU(2)$ singlets $p_{Rj}$
with hypercharge $2/3$.
The $n_{R1}$ and $p_{R1}$ are $A_4$-invariant
(they are $\bf{1}$'s of $A_4$),
the $n_{R2}$ and $p_{R2}$ are $\bf{1^\prime}$'s of $A_4$,
and the $n_{R3}$ and $p_{R3}$ are $\bf{1^{\prime\prime}}$'s of $A_4$.
This means that there exist
two non-commuting transformations $T_1$ and $T_2$,
\ba
T_1: & & \left\{ \begin{array}{l}
\phi_1 \to \phi_2 \to \phi_3 \to \phi_1,
\\
Q_{L1} \to Q_{L2} \to Q_{L3} \to Q_{L1},
\\
n_{R2} \to \omega n_{R2}, \ n_{R3} \to \omega^2 n_{R3},
\\
p_{R2} \to \omega p_{R2}, \ p_{R3} \to \omega^2 p_{R3},
\end{array} \right.
\label{t1} \\
T_2: & & \left\{ \begin{array}{l}
\phi_2 \to - \phi_2, \ \phi_3 \to - \phi_3,
\\
Q_{L2} \to - Q_{L2}, \ Q_{L3} \to - Q_{L3},
\end{array} \right.
\label{t2}
\ea
under which the Lagrangian is invariant.
In equation~(\ref{t1}),
$\omega \equiv \exp{\left( 2 i \pi / 3 \right)}
= \sqrt[3]{1}
= \left( - 1 + i\, \sqrt{3} \right) / 2$.

The $\bf{3}$ is a real representation of $A_4$.
Indeed,
the scalar $SU(2)$ doublets with hypercharge $-1/2$
\be
\tilde \phi_j \equiv i \tau_2 \phi_j^\ast
\ee
transform under $T_1$ and $T_2$ in exactly the same way as the $\phi_j$,
as is obvious from equations~(\ref{t1}) and~(\ref{t2}).

Given this spectrum of fields
and their transformation laws
under both the gauge symmetry $SU(2) \times U(1)$
and the flavour symmetry $A_4$,
the quark Yukawa Lagrangian is
\ba
{\cal L}_{\rm Yukawa} &=&
- y_1 \left( \overline{Q_{L1}} \phi_1
+ \overline{Q_{L2}} \phi_2
+ \overline{Q_{L3}} \phi_3
\right) n_{R1}
\no & &
- y_2 \left( \overline{Q_{L1}} \phi_1
+ \omega \overline{Q_{L2}} \phi_2
+ \omega^2 \overline{Q_{L3}} \phi_3
\right) n_{R2}
\no & &
- y_3 \left( \overline{Q_{L1}} \phi_1
+ \omega^2 \overline{Q_{L2}} \phi_2
+ \omega \overline{Q_{L3}} \phi_3
\right) n_{R3}
\no & &
- y_4 \left( \overline{Q_{L1}} \tilde \phi_1
+ \overline{Q_{L2}} \tilde \phi_2
+ \overline{Q_{L3}} \tilde \phi_3
\right) p_{R1}
\no & &
- y_5 \left( \overline{Q_{L1}} \tilde \phi_1
+ \omega \overline{Q_{L2}} \tilde \phi_2
+ \omega^2 \overline{Q_{L3}} \tilde \phi_3
\right) p_{R2}
\no & &
- y_6 \left( \overline{Q_{L1}} \tilde \phi_1
+ \omega^2 \overline{Q_{L2}} \tilde \phi_2
+ \omega \overline{Q_{L3}} \tilde \phi_3
\right) p_{R3}
+ {\rm H.c.},
\ea
the six Yukawa couplings $y_{1\mbox{--}6}$ being in general complex.

The scalar doublets
\be
\phi_j = \left( \begin{array}{c}
\phi_j^+ \\ \phi_j^0
\end{array} \right),
\quad
\tilde \phi_j = \left( \begin{array}{c}
{\phi_j^0}^\ast \\ - \phi_j^-
\end{array} \right)
\ee
are assumed to have vacuum expectation values (VEVs)
\be
\left\langle 0 \left| \phi_1^0 \right| 0 \right\rangle
= v_1 e^{- i \alpha / 2},
\quad
\left\langle 0 \left| \phi_2^0 \right| 0 \right\rangle
= v_2 e^{i \beta / 2},
\quad
\left\langle 0 \left| \phi_3^0 \right| 0 \right\rangle
= v_3,
\ee
where $v_{1,2,3}$ are,
without loss of generality,
real and non-negative.
Since
$\overline{Q_{Lj}} = \left( \overline{p_{Lj}},\ \overline{n_{Lj}} \right)$,
the quark mass matrices,
defined through
\be
{\cal L}_{\rm mass} =
- \left( \begin{array}{ccc}
\overline{n_{L1}} & \overline{n_{L2}} & \overline{n_{L3}}
\end{array} \right)
M_n
\left( \begin{array}{c}
n_{R1} \\ n_{R2} \\ n_{R3}
\end{array} \right)
-
\left( \begin{array}{ccc}
\overline{p_{L1}} & \overline{p_{L2}} & \overline{p_{L3}}
\end{array} \right)
M_p
\left( \begin{array}{c}
p_{R1} \\ p_{R2} \\ p_{R3}
\end{array} \right)
+ {\rm H.c.},
\ee
are
\ba
\label{M_n}
M_n &=& D
\left( \begin{array}{ccc}
y_1 v_1 & y_2 v_1 & y_3 v_1 \\
y_1 v_2 & \omega y_2 v_2 & \omega^2 y_3 v_2 \\
y_1 v_3 & \omega^2 y_2 v_3 & \omega y_3 v_3
\end{array} \right),
\\
\label{M_p}
M_p &=& D^\ast
\left( \begin{array}{ccc}
y_4 v_1 & y_5 v_1 & y_6 v_1 \\
y_4 v_2 & \omega y_5 v_2 & \omega^2 y_6 v_2 \\
y_4 v_3 & \omega^2 y_5 v_3 & \omega y_6 v_3
\end{array} \right),
\ea
where
\be
D \equiv \mbox{diag}
\left( e^{- i \alpha / 2},\ e^{i \beta / 2},\ 1 \right).
\ee

The quark mass matrices in equations~(\ref{M_n}) and~(\ref{M_p}) are identical
to the mass matrix for the charged leptons
in the original $A_4$ model of Ma and Rajasekaran (MR)~\cite{raja}.
Indeed,
we have adopted in our model
the same $A_4$ representations for the quarks
as MR did in their paper for the leptons.
The crucial difference between the two models is that,
while MR have assumed a vacuum state
characterized by $v_1 = v_2 = v_3$ and $\alpha = \beta = 0$,
we shall demonstrate the existence of,
and employ,
a vacuum state with different features.

Let the unitary matrices $U_{L,R}^{n,p}$ satisfy
\ba
{U_L^n}^\dagger
\left( \begin{array}{ccc}
y_1 v_1 & y_2 v_1 & y_3 v_1 \\
y_1 v_2 & \omega y_2 v_2 & \omega^2 y_3 v_2 \\
y_1 v_3 & \omega^2 y_2 v_3 & \omega y_3 v_3
\end{array} \right)
U_R^n &=& {\rm diag} \left( m_d, m_s, m_b \right),
\\
{U_L^p}^\dagger
\left( \begin{array}{ccc}
y_4 v_1 & y_5 v_1 & y_6 v_1 \\
y_4 v_2 & \omega y_5 v_2 & \omega^2 y_6 v_2 \\
y_4 v_3 & \omega^2 y_5 v_3 & \omega y_6 v_3
\end{array} \right)
U_R^p &=& {\rm diag} \left( m_u, m_c, m_t \right).
\ea
Then,
the quark mixing (CKM) matrix is
\be
V =
{U_L^p}^\dagger D^2\, U_L^n.
\ee

One may absorb the phases of $y_{1,2,3}$ in the overall phases
of the three rows of $U_R^n$,
and similarly absorb the phases of $y_{4,5,6}$ in the matrix $U_R^p$.
Those six phases are therefore unphysical.
Thus,
this model for the quark masses and mixings has ten parameters:
the two phases $\alpha$ and $\beta$ in the diagonal matrix $D^2$,
and the eight real quantities
\be
|y_1| v_3,\
\left| \frac{y_2}{y_1} \right|,\
\left| \frac{y_3}{y_1} \right|,\
|y_4| v_3,\
\left| \frac{y_5}{y_4} \right|,\
\left| \frac{y_6}{y_4} \right|,\
\frac{v_2}{v_3},\
\frac{v_1}{v_3}.
\ee
As we shall see in the next section,
the $A_4$-symmetric scalar potential is so constrained
that these ten parameters reduce to only eight.

\section{The scalar potential}
\label{potential}

The most general renormalizable scalar potential
invariant under the symmetry $A_4$ is
\ba
V &=& \mu \left( \phi_1^\dagger \phi_1
+ \phi_2^\dagger \phi_2 + \phi_3^\dagger \phi_3 \right)
+ \lambda_1 \left( \phi_1^\dagger \phi_1
+ \phi_2^\dagger \phi_2 + \phi_3^\dagger \phi_3 \right)^2
\no & &
+ \lambda_2 \left[
\left( \phi_1^\dagger \phi_1 \right)
\left( \phi_2^\dagger \phi_2 \right)
+ \left( \phi_2^\dagger \phi_2 \right)
\left( \phi_3^\dagger \phi_3 \right)
+ \left( \phi_3^\dagger \phi_3 \right)
\left( \phi_1^\dagger \phi_1 \right)
\right]
\no & &
+ \left( \lambda_3 - \lambda_2 \right) \left[
\left( \phi_1^\dagger \phi_2 \right)
\left( \phi_2^\dagger \phi_1 \right)
+ \left( \phi_2^\dagger \phi_3 \right)
\left( \phi_3^\dagger \phi_2 \right)
+ \left( \phi_3^\dagger \phi_1 \right)
\left( \phi_1^\dagger \phi_3 \right)
\right]
\no & &
+ \frac{\lambda_4}{2} \left\{
e^{i \epsilon} \left[
\left( \phi_1^\dagger \phi_2 \right)^2
+ \left( \phi_2^\dagger \phi_3 \right)^2
+ \left( \phi_3^\dagger \phi_1 \right)^2 \right]
+ \mbox{H.c.} \right\},
\label{thepot}
\ea
where $\mu$ and $\lambda_{1\mbox{--}4}$ are real.
The phase $\epsilon$ is arbitrary.

We define $v \equiv \sqrt{v_1^2 + v_2^2 + v_3^2}$ and
\ba
\theta_1 &\equiv& \epsilon - \beta,
\\
\theta_2 &\equiv& \epsilon - \alpha,
\\
\theta_3 &\equiv& \epsilon + \alpha + \beta.
\ea
Then,
\ba
V_0 \equiv \left\langle 0 \left| V \right| 0 \right\rangle &=&
\mu v^2 + \lambda_1 v^4
+ \lambda_3 \left( v_1^2 v_2^2 + v_2^2 v_3^2 + v_3^2 v_1^2 \right)
\no & &
+ \lambda_4 \left(
v_1^2 v_2^2 \cos{\theta_3}
+ v_2^2 v_3^2 \cos{\theta_1}
+ v_3^2 v_1^2 \cos{\theta_2} \right).
\ea
The equations for vacuum stability are
\ba
\frac{\partial V_0}{\partial v_1^2}
= 0 &=&
\mu + 2 \lambda_1 v^2 + \lambda_3 \left( v_2^2 + v_3^2 \right)
+ \lambda_4 \left( v_2^2 \cos{\theta_3} + v_3^2 \cos{\theta_2} \right),
\label{v1} \\
\frac{\partial V_0}{\partial v_2^2}
= 0 &=&
\mu + 2 \lambda_1 v^2 + \lambda_3 \left( v_1^2 + v_3^2 \right)
+ \lambda_4 \left( v_1^2 \cos{\theta_3} + v_3^2 \cos{\theta_1} \right),
\label{v2} \\
\frac{\partial V_0}{\partial v_3^2}
= 0 &=&
\mu + 2 \lambda_1 v^2 + \lambda_3 \left( v_1^2 + v_2^2 \right)
+ \lambda_4 \left( v_1^2 \cos{\theta_2} + v_2^2 \cos{\theta_1} \right),
\label{v3} \\
\frac{\partial V_0}{\partial \alpha}
= 0 &=& \lambda_4 v_1^2 \left(
- v_2^2 \sin{\theta_3} + v_3^2 \sin{\theta_2} \right),
\label{alpha} \\
\frac{\partial V_0}{\partial \beta}
= 0 &=& \lambda_4 v_2^2 \left(
- v_1^2 \sin{\theta_3} + v_3^2 \sin{\theta_1} \right).
\label{beta}
\ea
We reject possible solutions to these equations
in which one of the VEVs vanishes,
and also solutions in which $v_1 = v_2 = v_3$.\footnote{Notice that
Ma and Rajasekaran opted precisely for $v_1 = v_2 = v_3$
in their model~\cite{raja},
which is otherwise similar to ours.}
Then,
equations~(\ref{alpha}) and~(\ref{beta}) yield
\be
\sin{\theta_j} = k v_j^2,
\label{sines}
\ee
where $k$ is a real constant with dimension $M^{-2}$.
Subtracting equations~(\ref{v2}) and~(\ref{v3}) from equation~(\ref{v1}),
one obtains
\be
\begin{array}{rcl}
\left( v_2^2 - v_1^2 \right) \lambda_3
+ \left[ \left( v_2^2 - v_1^2 \right) \cos{\theta_3}
+ v_3^2 \left( \cos{\theta_2} - \cos{\theta_1} \right) \right] \lambda_4
&=& 0,
\\*[2mm]
\left( v_3^2 - v_1^2 \right) \lambda_3
+ \left[ \left( v_3^2 - v_1^2 \right) \cos{\theta_2}
+ v_2^2 \left( \cos{\theta_3} - \cos{\theta_1} \right) \right] \lambda_4
&=& 0.
\end{array}
\label{syst}
\ee
Equations~(\ref{syst})
constitute a Cramer system for $\lambda_3$ and $\lambda_4$.
The Cramer determinant must vanish and one hence obtains
\be
\sum_{j=1}^3 a_j \cos{\theta_j} = 0,
\label{sum}
\ee
where
\ba
a_1 &\equiv& v_3^4 - v_2^4 + v_1^2 v_2^2 - v_1^2 v_3^2,
\label{a1} \\
a_2 &\equiv& v_1^4 - v_3^4 + v_2^2 v_3^2 - v_1^2 v_2^2,
\label{a2} \\
a_3 &\equiv& v_2^4 - v_1^4 + v_1^2 v_3^2 - v_2^2 v_3^2
\label{a3}
\ea
satisfy
\be
\sum_{j=1}^3 a_j = 0.
\label{sum2}
\ee
Equation~(\ref{sum}) together with equation~(\ref{sines}) imply
\ba
0 &=&
- \lambda \left( a_1 \sqrt{1 - k^2 v_1^4},\
a_2 \sqrt{1 - k^2 v_2^4},\ a_3 \sqrt{1 - k^2 v_3^4} \right)
\no &=&
4 k^2
\left[ 4 k^2 v_1^4 v_2^4 v_3^4 - \lambda \left( v_1^2, v_2^2, v_3^2 \right)
\right]
\left( v_1^2 - v_2^2 \right)^2
\left( v_1^2 - v_3^2 \right)^2
\left( v_2^2 - v_3^2 \right)^2,
\label{fin}
\ea
where~\cite{book}
\be
\lambda \left( a, b, c \right) \equiv
- a^4 - b^4 - c^4 + 2 a^2 b^2 + 2 a^2 c^2 + 2 b^2 c^2.
\ee
Equation~(\ref{fin}) has five solutions:
\begin{enumerate}
\item $k = 0$,
\item $k^2 = \lambda \left( v_1^2, v_2^2, v_3^2 \right)
/ \left( 4 v_1^4 v_2^4 v_3^4 \right)$,
\item $v_1 = v_2$,
\item $v_1 = v_3$,
\item $v_2 = v_3$.
\end{enumerate}
It is easy to explore in detail solutions 1\ and~2\ and to show that
they require $\lambda_3 = \pm \lambda_4$
(solution 1\ furthermore needs $\epsilon = 0$).
Thus,
those solutions require a non-trivial constraint
on the parameters of the potential,
and that constraint is in general unstable under renormalization.
Those solutions should therefore be discarded.
The remaining solutions 3--5 are equivalent,
since the three scalar doublets $\phi_j$ form a triplet of $A_4$.
We shall use for definiteness solution 3:
$v_1 = v_2$ and $\theta_1 = \theta_2$,
i.e.~$\alpha = \beta$.
Equations~(\ref{v1})--(\ref{beta}) then reduce to only three equations,
\ba
0 &=&
\mu + 2 \lambda_1 \left( 2 v_1^2 + v_3^2 \right)
+ \lambda_3 \left( v_1^2 + v_3^2 \right)
+ \lambda_4 \left[ v_1^2 \cos{\left( \epsilon + 2 \alpha \right)}
+ v_3^2 \cos{\left( \epsilon - \alpha \right)} \right],
\\
0 &=&
\mu + 2 \lambda_1 \left( 2 v_1^2 + v_3^2 \right)
+ 2 \lambda_3 v_1^2 + 2 \lambda_4 v_1^2 \cos{\left( \epsilon - \alpha \right)},
\\
0 &=& v_1^2 \sin{\left( \epsilon + 2 \alpha \right)}
- v_3^2 \sin{\left( \epsilon - \alpha \right)},
\ea
which determine the three quantities $v_1$,
$v_3$,
and $\alpha$.

\section{The mass matrices and $CP$ conservation}
\label{matrices}

We saw in the previous section that,
just as we had advertised at the end of section~\ref{yukawas},
consideration of the most general $A_4$-invariant scalar potential
actually reduces the ten parameters of our model to only eight,
since $v_2 = v_1$ and $\beta = \alpha$.
The quark mass matrices of our model are therefore
\ba
M_n &=&
\mbox{diag} \left( e^{- i \alpha / 2}, e^{i \alpha / 2}, 1 \right)
\left( \begin{array}{ccc}
a & b & c \\
a & \omega b & \omega^2 c \\
r a & \omega^2 r b & \omega r c
\end{array} \right),
\label{mn} \\
M_p &=&
\mbox{diag} \left( e^{i \alpha / 2}, e^{- i \alpha / 2}, 1 \right)
\left( \begin{array}{ccc}
f & g & h \\
f & \omega g & \omega^2 h \\
r f & \omega^2 r g & \omega r h
\end{array} \right),
\label{mp}
\ea
where $r \equiv v_3 / v_1$ and $a$,
$b$,
$c$,
$f$,
$g$,
and $h$ are real and positive.
Then,
\ba
H_n \equiv M_n M_n^\dagger &=&
\left( \begin{array}{ccc}
x & y^\ast e^{- i \alpha} & r y e^{- i \alpha / 2} \\
y e^{i \alpha} & x & r y^\ast e^{i \alpha / 2} \\
r y^\ast e^{i \alpha / 2} & r y e^{- i \alpha / 2} & r^2 x
\end{array} \right),
\\
H_p \equiv M_p M_p^\dagger &=&
\left( \begin{array}{ccc}
z & w^\ast e^{i \alpha} & r w e^{i \alpha / 2} \\
w e^{- i \alpha} & z & r w^\ast e^{- i \alpha / 2} \\
r w^\ast e^{- i \alpha / 2} & r w e^{i \alpha / 2} & r^2 z
\end{array} \right),
\ea
where
$x \equiv a^2 + b^2 + c^2$ and $z \equiv f^2 + g^2 + h^2$ are real,
while
$y \equiv a^2 + \omega b^2 + \omega^2 c^2$
and $w \equiv f^2 + \omega g^2 + \omega^2 h^2$
are complex.
Now,
computing the commutator of $H_p$ and $H_n$ one finds that it is of the form
\be
\left[ H_p, H_n \right] = \left( \begin{array}{ccc}
- n & 0 & - m \\ 0 & n & - m^\ast \\ m^\ast & m & 0
\end{array} \right),
\ee
hence $\det \left[ H_p, H_n \right] = 0$.
Therefore,
in this model there is no $CP$ violation in the quark mixing matrix,
i.e.~the Jarlskog observable $J$~\cite{jarlskog} vanishes.

\section{Numerical procedure and results}
\label{chi2}

We have performed a global $\chi^2$~analysis
of the the quark mass matrices given
in the previous section---equations~(\ref{mn}) and~(\ref{mp})---by
employing the downhill simplex method~\cite{downhill}.
Table~\ref{input} specifies in its first two columns
the observable quantities $O_i$ in the form
\be
O_i = \bar{O}_i \pm \sigma_i,
\ee
where $\bar O_i$ is the experimental mean value of $O_i$
and $\sigma_i$ is the square root of its variance.
The index $i = 1, \ldots, 9$ labels the nine observables
given in table~\ref{input}.
\begin{table}[t]
\begin{center}
\begin{tabular}{c|c|c|c}
\hline\hline Observable & Experimental value & Model prediction &
Pull  \\ \hline
$m_{d}$~[MeV] & $5 \pm 2$ & $4.977$ & $-1.2 \times 10^{-2}$ \\
$m_{s}$~[MeV] & $95 \pm 25$ & $90.545$ & $-1.8 \times 10^{-1}$ \\
$m_{b}$~[MeV] & $4200 \pm 70$ & $4200.79$ & $+1.1 \times 10^{-2}$ \\[1ex]
$m_{u}$~[MeV] & $2.25 \pm 0.75$  & $2.250$ & $+1.9 \times 10^{-5}$ \\
$m_{c}$~[MeV] & $1250 \pm 90$  & $1250.498$ & $+5.5 \times 10^{-3}$ \\
$m_{t}$~[GeV] & $172.5 \pm 2.7$ & $172.497$ & $-1.2 \times 10^{-3}$ \\[1ex]
$\sin \theta_{12}$ & $0.2243 \pm 0.0016$  & $0.22431$ &
$+8.1 \times 10^{-3}$ \\
$\sin \theta_{23}$ & $0.0413 \pm 0.0015$  & $0.04139$ &
$+6.0 \times 10^{-2}$ \\
$\sin \theta_{13}$ & $0.0037 \pm 0.0005$ & $0.003627$ &
$-1.5 \times 10^{-1}$ \\
\hline\hline
\end{tabular}
\end{center}
\caption{
Experimental data and result of our best fit.
The experimental data
(average values and error bars)
used in our numerical analysis are given in the second column.
The data on the quark masses have been taken from~\cite{masses}.
The light-quark masses are
renormalized at a common scale $\mu \approx$~2~GeV.
The charm- and bottom-quark masses are running masses
in the $\overline{\mathrm{MS}}$ scheme.
The data on the quark mixing angles
have been taken from~\cite{mixingangles}.
The third column displays the values $P_i$ predicted by our model
when the values of its parameters are those in equations~(\ref{param}).
The fourth column
shows the number of
standard deviations from the mean values,
$\left( P_i - \bar{O}_i \right) / \sigma_i$,
computed using the data from the second column.
The value $\chi^2 = 0.057$ is the sum of the squares
of the numbers in the fourth
column and is dominated by the pulls of $m_s$ and $\sin{\theta_{13}}$.}
\label{input}
\end{table}
Writing $\mathbf{x}$ for the set of the eight parameters
of our model
($a$,
$b$,
$c$,
$f$,
$g$,
$h$,
$r$,
and $\alpha$),
and $P_i \left( \mathbf{x} \right)$
for the resulting predictions for each of the observables,
one constructs the $\chi^{2}$ function
\be
\label{chisquare}
\chi^{2} \left( \mathbf{x} \right)
= \sum_{i=1}^{9}
\left[ \frac{P_i \left( \mathbf{x} \right) - \bar{O}_i}
{\sigma_i} \right]^2.
\ee
The global minimum of $\chi^2$ represents the best possible fit
of the model predictions to the experimental data.

We found an excellent fit,
with $\chi^2=0.057$,
of our model to the nine input data specified in table~\ref{input}.
The input parameters of the fit are
\be
\begin{array}{rcl}
a &=& 40.75189\,\textrm{MeV}, \\
b &=& 87.78761\,\textrm{MeV}, \\
c &=& 2.347665\,\textrm{MeV}, \\
f &=& 3941.127\,\textrm{MeV}, \\
g &=& 515.0460\,\textrm{MeV}, \\
h &=& 1.060808\,\textrm{MeV}, \\
r &=& 43.37746, \\
\alpha &=& 0.2251660\ \textrm{radians}.
\end{array}
\label{param}
\ee
Other details of the fit are given in the third and fourth columns
of table~\ref{input}.

Notice that the observables in table~\ref{input} do not include
the phase $\delta$ of the quark mixing matrix.
Indeed,
since we already know that in our model that matrix is real,
we are of course unable to fit for any $\delta \neq 0, \pi$.
In our model the explanation of all the observed $CP$ violation
through a sole phase in the quark mixing matrix gets spoiled:
other sources of $CP$ violation must be found,
and fitted to the observed data.
In our model the most likely source of $CP$ violation
will be the interactions mediated by both charged and neutral scalars,
since ours is a three-Higgs-doublet model featuring,
in particular,
flavour-changing interactions mediated by neutral scalars.

In order to test the variation of $\chi^2$
as a function of the value $\widehat O_i$ of an observable quantity
$O_i$,
we substitute in the expression
for $\chi^2 \left( \mathbf{x} \right)$
the term
$\left[ P_i \left( \mathbf{x} \right) - \bar{O}_i \right]^2
/ \left( \sigma_i \right)^2$
by a term
$\left[ P_i \left( \mathbf{x} \right) - \widehat O_i \right]^2
/ \left( 0.01\ \widehat O_i \right)^2$.
The small error assigned to $\widehat O_i$
in the denominator of this term
guarantees that
$O_i$ gets pinned down to the value $\widehat O_i$.

\begin{figure}[t]
\begin{center}
\epsfig{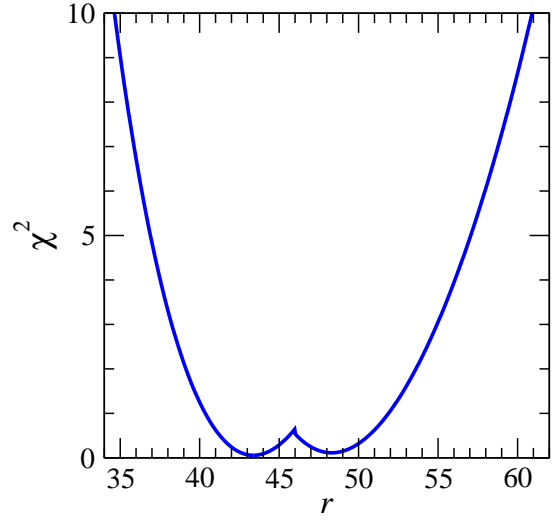}
\end{center}
\caption{$\chi^2$ as a function of the ratio of VEVs $r \equiv v_3 / v_1$.
\label{fig_r}}
\end{figure}

Figure~\ref{fig_r} depicts $\chi^2$ as a function of $r$,
i.e.~of the ratio of VEVs $v_3/v_1$.
We read off from that figure that
only for $40 \lesssim r \lesssim 52$ can good fits be obtained;
thus,
the range of the ratio of VEVs is severely constrained.

\begin{figure}[b]
\begin{center}
\epsfig{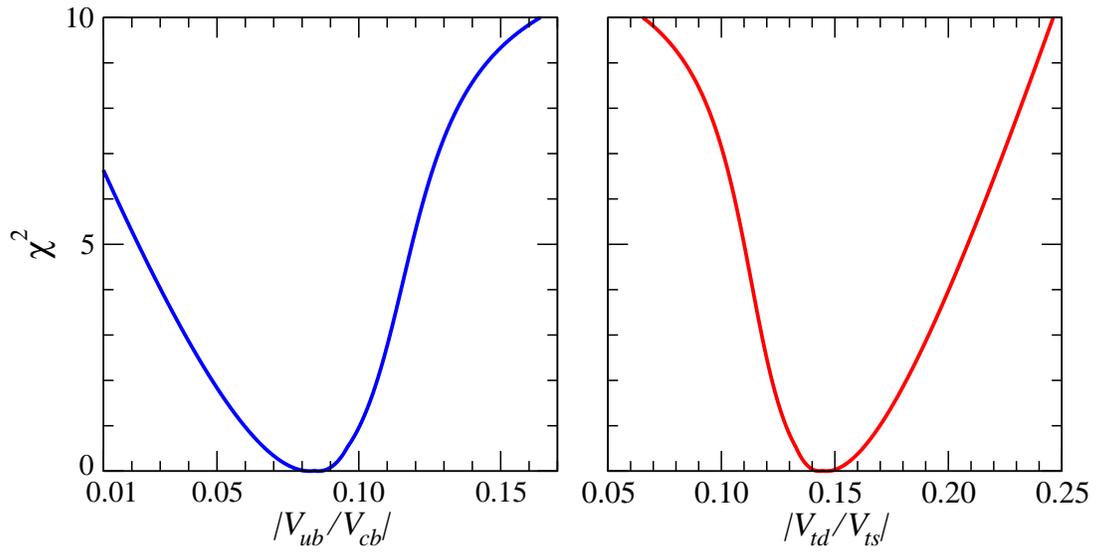}
\end{center}
\caption{$\chi^2$ as a function of the CKM-matrix parameters
$\left|V_{ub} / V_{cb}\right|$ and $\left|V_{td} / V_{ts}\right|$.
\label{fig_VubVcb}}
\end{figure}

In figure~\ref{fig_VubVcb}~(left panel),
the change of $\chi^2$ under variations of the quark-mixing observable
$\left|V_{ub} / V_{cb}\right|$ is shown.
There is a pronounced minimum of $\chi^2$
for $0.08 \lesssim \left| V_{ub} / V_{cb} \right| \lesssim 0.09$;
this is in excellent agreement
with the value obtained for that observable by the phenomenological analyses.
This remarkable result of our model
provides a clear-cut prediction for $\left|V_{ub} / V_{cb}\right|$.
Figure~\ref{fig_VubVcb}~(right panel) gives $\chi^2$
as a function of $\left| V_{td} / V_{ts} \right|$.
We find in this case excellent fits whenever
$0.14 \lesssim \left| V_{td} / V_{ts} \right| \lesssim 0.15$.
Clearly,
this result is correlated to our model's prediction
for $\left|V_{ub} / V_{cb}\right|$,
since in our model $CP$ is conserved in quark mixing
and therefore the CKM matrix is determined by only three parameters.

\section{Conclusions}
\label{summary}

In this paper we have proposed a self-consistent model
for the quark masses and mixings based on a family symmetry $A_4$.
The Yukawa-coupling matrices of our model contain,
at face value,
ten parameters,
but,
when one considers the $A_4$-symmetric scalar potential in detail,
one sees that two of those parameters actually disappear.
In our model the family symmetry $A_4$ is not broken
anywhere in the Lagrangian---its breaking is fully spontaneous.
The model gives a perfect fit of the observed quark masses
and mixing parameters,
except for the fact that
there is no $CP$ violation at all in the CKM matrix.
The observed $CP$ violation should result in our model
from scalar-mediated interactions,
in particular flavour-changing neutral Yukawa interactions
at tree level
and also charged-scalar-mediated box diagrams at the one-loop level;
a detailed study of those interactions should be the subject
of a separate publication.

In previous $A_4$ models by other authors
a vacuum characterized by equal VEVs $v_1 = v_2 = v_3$
has often been employed.
In our $A_4$ model for the quark sector,
on the contrary,
we use $v_1 = v_2$ while $v_3$ is some 43 times larger.
It is not clear to us whether this vacuum may or may not produce
a viable model for the lepton sector as well;
this will be the subject of future investigation.

\paragraph{Acknowledgement:}
The work of L.L.\ was supported by the Portuguese
\textit{Funda\c c\~ao para a Ci\^encia e a Tecnologia}
through the project U777--Plurianual.

\newpage

\begin{thebibliography}{9}

\bibitem{zeros}
W.~Grimus, A.S.~Joshipura, L.~Lavoura, and M.~Tanimoto,
\textit{Symmetry realization of texture zeros},
Eur.~Phys.~J.~C \textbf{36} (2004) 227
[hep-ph/0405016].

\bibitem{kephart}
P.H.~Frampton and T.W.~Kephart,
\textit{Classification of conformality models
based on non-Abelian orbifolds},
Phys.~Rev.~D \textbf{64} (2001) 086007
[hep-ph/0011186].

\bibitem{leptons}
See E.~Ma,
\textit{$A_4$ symmetry and neutrinos}
[arXiv:0710.3851],
and the references therein.

\bibitem{quarks}
E.~Ma,
\textit{Quark mass matrices in the $A_4$ model},
Mod.~Phys.~Lett.~A \textbf{17} (2002) 627
[hep-ph/0203238];
E.~Ma, H.~Sawanaka, and M.~Tanimoto,
\textit{Quark masses and mixing with $A_4$ family symmetry},
Phys.~Lett.~B \textbf{641} (2006) 301
[hep-ph/0606103].

\bibitem{both}
K.S.~Babu, E.~Ma, and J.W.F.~Valle,
\textit{Underlying $A_4$ symmetry for the neutrino mass matrix
and the quark mixing matrix},
Phys.~Lett.~B \textbf{552} (2003) 207
[hep-ph/0206292];
Xiao-Gang~He, Yong-Yeon~Keum, and R.R.~Volkas,
\textit{$A_4$ flavour symmetry breaking scheme
for understanding quark and neutrino mixing angles},
JHEP \textbf{0604} (2006) 039
[hep-ph/0601001];
F.~Bazzocchi, S.~Kaneko, and S.~Morisi,
\textit{A SUSY $A_4$ model for fermion masses and mixings}
[arXiv:0707.3032].

\bibitem{raja}
E.~Ma and G.~Rajasekaran,
\textit{Softly broken $A_4$ symmetry
for nearly degenerate neutrino masses},
Phys. Rev. D \textbf{64} (2001) 113012 [hep-ph/0106291].

\bibitem{book}
G.C.~Branco, L.~Lavoura, and J.P.~Silva,
\textit{CP violation}
(Oxford University Press, 1999),
p.~167.

\bibitem{jarlskog}
C.~Jarlskog,
\textit{Commutator of the quark mass matrices
in the standard electroweak model
and a measure of maximal $CP$ violation},
Phys.~Rev.~Lett.\ \textbf{55} (1985) 1039;
\textit{A basis-independent formulation of the connection
between quark mass matrices,
$CP$ violation,
and experiment},
Z.~Phys.~C \textbf{29} (1985) 491.
See also J.~Bernab\'eu, G.C.~Branco, and M.~Gronau,
\textit{$CP$ restrictions on quark mass matrices},
Phys.~Lett.~B \textbf{169} (1986) 243.

\bibitem{downhill}
J.A.~Nelder and R.~Mead,
\textit{A simplex method for function minimization},
Comput.~J.~\textbf{7} (1965) 308;
W.H.~Press, B.P.~Flannery, S.A.~Teukolsky, and W.T.~Vetterling,
\textit{Numerical recipes in C: The art of scientific computing}
(Cambridge University Press, 1992).

\bibitem{masses}
W.-M.~Yao \textit{et al.},
\textit{Review of particle physics},
J.~Physics~G \textbf{33} (2006) 1.

\bibitem{mixingangles}
S.~Eidelman \textit{et al.},
\textit{Review of particle physics},
Phys.~Lett.~B \textbf{592} (2004) 1.

\end{thebibliography}
\end{document}